\documentclass{article}

\usepackage{PRIMEarxiv}

\usepackage[utf8]{inputenc} 
\usepackage[T1]{fontenc}    
\usepackage{hyperref}       
\usepackage{url}            
\usepackage{booktabs}       
\usepackage{amsfonts}       
\usepackage{nicefrac}       
\usepackage{microtype}      
\usepackage{lipsum}
\usepackage{fancyhdr}       
\usepackage{graphicx}       
\usepackage{xcolor}
\usepackage{tabularx}
\usepackage{amsmath}
\usepackage{physics}
\usepackage{float}
\DeclareRobustCommand{\bigO}{%
  \text{\usefont{OMS}{cmsy}{m}{n}O}%
} 
\title{Removing constraints of 4D-STEM with a framework for event-driven acquisition and processing}

\author{
  Arno Annys$^{1,2,*}$, Hoelen L. Lalandec Robert$^{1,2}$, Saleh Gholam$^{1,2}$, Joke Hadermann$^{1,2}$, Jo Verbeeck$^{1,2}$ \\
  $^{1}$Electron Microscopy for Materials Science (EMAT), University of Antwerp, Groenenborgerlaan 171, 2020 Antwerp, Belgium
  \\
$^{2}$NANOlight Center of Excellence, University of Antwerp, Groenenborgerlaan 171, 2020 Antwerp, Belgium
\\$^{*}$Corresponding author: arno.annys@uantwerpen.be 
}

\begin{document}
This is the author's accepted manuscript (AAM), post peer review. The final published version is available at: \href{https://doi.org/10.1016/j.ultramic.2025.114206}{10.1016/j.ultramic.2025.114206}
\\
\\
\maketitle

\begin{abstract}
Pixelated detectors in scanning transmission electron microscopy (STEM) generate large volumes of data, often tens to hundreds of GB per scan. However, to make current advancements scalable and enable widespread adoption, it is essential to use the most efficient representation of an electron's information. Event-driven direct electron detectors, such as those based on the Timepix3 chip, offer significant potential for electron microscopy, particularly for low-dose experiments and real-time data processing. In this study, we compare sparse and dense data representations in terms of their size and computational requirements across various 4D-STEM scenarios, including high-resolution imaging and nano-beam electron diffraction. The advantages of performing 4D-STEM in an event-driven mode—such as reduced requirements in memory, bandwidth, and computational demands—can only be fully leveraged if the entire acquisition and processing pipeline is optimized to work directly with the event format, avoiding intermediate dense representations. We introduce a framework designed for acquisition and processing based on this event format, and demonstrate live processing of event-driven 4D-STEM, including analytical ptychography.
\end{abstract}


\section*{Introduction}
    
    Annular dark field scanning transmission electron microscopy (ADF-STEM) has long been one of the most popular tools in electron microscopy for the investigation of materials, especially at high spatial resolution \cite{Pennycook2012}. One of the interests of ADF-STEM is that it provides interpretable high-resolution images without the need for extensive processing or reference simulations, unlike many conventional transmission electron microscopy (CTEM) methods. One of the disadvantages of ADF-STEM is that only a small part of the transmitted electrons is used, limiting its information content and dose-efficiency, and thus making it a rather poor choice for investigating beam-sensitive materials such as biological objects. Thanks to advances in pixelated electron detectors, especially direct electron detectors (DED) \cite{McMullan2007,Llopart2007,Ballabriga2011,Plackett2013,Poikela2014,Ryll2016,Tate2016,Philipp2022,Llopart2022,Zambon2023,Ercius2024}, a more powerful tool has emerged, where for each probe position of a scan, the diffracted intensity is recorded \cite{Muller2012a}. This results in two-dimensional diffraction patterns measured for a two-dimensional series of probe positions, yielding a four-dimensional dataset. Therefore, this technique is usually dubbed 4D-STEM \cite{Yang2015a} or momentum-resolved STEM \cite{Muller-Caspary2018a} by the community. 4D-STEM allows one to apply multiple information retrieval techniques with different characteristics from the same dataset. In nano-beam electron diffraction, 4D-STEM provides a powerful and versatile tool for e.g. strain, orientation and crystal structure mapping \cite{Uesugi2011,Beche2009,Wu2023b,panova2016,Kobler2013,Gallagher-Jones2020,Brunetti2011,Viladot2013}. For high-resolution microscopy, one of the most powerful methods is ptychography \cite{Rodenburg2019}, which is a class of computational imaging enabled by the coherent scattering of probing electrons by the electrostatic potential of the specimen. In particular, the analytical variants of the technique encompass the so-called sideband integration (SBI) \cite{Rodenburg1993,Pennycook2015,Yang2015b,Yang2016a,LalandecRobert2025} and Wigner distribution deconvolution (WDD) \cite{Rodenburg1992,McCallum1992,Nellist1994,Li2014} methods.
    
    Although 4D-STEM provides an effective characterization tool in a broad range of applications, it remains subject to a number of practical constraints. One of the most prominent ones is the large amount of raw data generated, and the computational efforts required to retrieve interpretable information from these large datasets. As a result, much of 4D-STEM data processing remains an afterthought, decoupled from the experiment. This means that the information richness is usually not leveraged by live processing and thus not driving the actual experiment. Another practical constraint is the frame rate limitation of most current-generation DED, also limiting live workflows. This work discusses possibilities to alleviate these constraints through sparse 4D-STEM workflows, mainly achieved by event-driven detection.

\section*{Direct electron detectors for STEM}
    
    DED are typically divided into two categories: monolithic active pixel sensors (MAPS) and hybrid pixel array detectors (HPAD). A thorough review of the characteristics of those two categories and the implications for electron microscopy is provided in reference \cite{Levin2021}. Typical characteristics of MAPS are small pixel pitches (often less than  10 $\mu$m), large numbers of pixels (several thousand squared), and frame rates in the kHz regime, making them well suited for CTEM. For STEM, HPAD are typically preferable, because they are made with a thick sensor layer that is separated from the electronics, which makes them radiation-hard, hence able to withstand localized beam intensity typical in a diffraction pattern. Due to the spread of the charges in the sensor layer and the need for pixel-wise electronics, the pixel pitches of HPAD must be larger, in the tens to hundreds of $\mu$m regime, and their number is limited to usually a few hundred squared. The frame rates of HPAD are typically higher than those of MAPS, having now surpassed the 100 kHz regime \cite{Zambon2023}.
    
    Upon electron incidence in an HPAD sensor, there is generally a cluster consisting of multiple neighboring pixels activated by a single electron \cite{Mir2017,Paton2021,Mangan2023}. The size of such a cluster depends on the sensor layer thickness, the sensor material, the energy deposited by the electron (the acceleration voltage in the case of full absorption) and the discriminator threshold of the detector.
    
    In terms of read-out mechanism, most HPAD like e.g. those based on the Medipix3 chip \cite{Ballabriga2011,Ballabriga2018} are still very similar to the CCD and CMOS detectors that have been used routinely in TEM. This read-out mechanism is typically called frame-based, where a value for every pixel is recorded after a chosen exposure period. The maximum frame rate of current generation frame-based HPAD is in the order of 100 kHz \cite{Zambon2023}, making them still at least an order of magnitude slower than what is frequently used for single-pixel detectors like in ADF-STEM. When no fast electrostatic beam blanker is available (synchronized with the scanning), this minimum frame time puts a lower limit on the electron dose for a scan with a workable fixed current and number of scan positions.
    
    Importantly, further improvements of the frame time in those DED will amplify a related issue. Frame-based detectors provide a fixed number of bytes for each frame, i.e. each probe position in a STEM scan. This means that pixels that were not activated by an electron still take up part of the bandwidth of the readout system of the detector, the bandwidth of the processing computer, and file storage space, without providing any information in exchange. This issue is especially pronounced in 4D-STEM for high-resolution imaging where, thanks to the latest advances in analysis methods like center of mass (CoM) \cite{Muller2014} or ptychography, sufficient information can already be obtained with just a few electron-hits in a single diffraction pattern \cite{OLeary2020,LalandecRobert2025}.
    
    Altering the read-out mechanism of a HPAD to an event-driven mechanism can drastically change the characteristics of the detector, even when the sensor layer is unchanged. In an event-driven read-out mechanism, each individual activated pixel (an event or hit) is immediately processed and transmitted as a data packet containing information such as position, timestamp, and energy, enabling continuous readout. The use of an event-driven HPAD in the electron microscope was already demonstrated in 1998 by Fan et al. \cite{Fan1998}, though limited in throughput and resolution by the technological constraints of the time. Thanks to advances in semiconductor technology, the Timepix3 chip \cite{Poikela2014,Frojdh2015} could be introduced, being the event-driven counterpart of the Medipix3 chip. The Timepix3 chip has a sparse and continuous read-out that provides time-of-arrival (TOA) and time-over-threshold (TOT) information with a nominal time resolution of 1.5625 ns. This has enabled the demonstration of several time-resolved \cite{Castioni2025,Haindl2023,Borrelli2024} and coincidence measurement methodologies \cite{Jannis2021,Varkentina2022,Preimesberger2025} in the electron microscope. The use of DED based on the Timepix3 has been demonstrated to allow 4D-STEM \cite{Jannis2022} and STEM electron energy loss spectroscopy (EELS) \cite{Auad2022} with sub-$\mu$s dwell times, allowing to match the scanning rates to those commonly used with single-pixel detectors like in ADF-STEM.
    
    The cryo-EM and microED community have also realized workflows exploiting the sparsity of electron detection, for example using a Timepix3-based DED \cite{VanSchayck2020,VanSchayck2023} or using the electron event representation (EER) as an in-line compression method \cite{Datta2021,Guo2020,Vlahakis2025}. In the case of the EER, the readout mechanism of the detector is not directly event-driven as in the case of Timepix3 and no TOA is available. Instead, event extraction from raw frames can be performed on a hardware level before crossing the bandwidth-limited communication channel to the user's PC. Localization of the electron impact with sub-pixel precision can be performed from the charge distribution if the frame is sufficiently sparse and coincidence losses can be avoided. EER is implemented on for example the Thermo Fisher Scientific Falcon 4i DED and the Direct Electron Apollo DED \cite{Peng2023}.

\section*{4D-STEM using a Timepix3}
    \label{sec:4DinTimepix}
    The procedure of performing 4D-STEM using an event-driven detector like a Timepix3-based DED was previously described in reference \cite{Jannis2022}. We provide a brief summary of the method and further expand on the key concepts such as synchronization, saturation and processing.
    \subsection*{Synchronization}
    As part of the 4D-STEM experiment, incident electrons are recorded continuously through activated pixels. An activated pixel is read out into a pair of pixel coordinates, the TOA and TOT. The coordinate of the electron on the detector directly provides the momentum information $(k_x,k_y)$, whereas the TOA encodes the real-space coordinates $(r_x,r_y)$ of the beam, through a synchronization with the scanning system. There are two main methods for synchronization: the use of a master clock for both the scan engine and the detector \cite{Jannis2022} or the use of triggers at each pixel or line \cite{Auad2022}. One could be tempted to ignore the need for an explicit synchronization between the detector and scan engine, but since the clock drift of crystal oscillator clocks is temperature dependent and in the parts per million regime \cite{Walls1995}, even fast measurements will be affected by clock drift. In general, both methods are suitable, although there are certain advantages associated with the use of a master clock. Most importantly, master clock synchronization allows each event to be treated completely independently, allowing a straightforward parallelization of the processing pipeline. This is not the case when triggers are used. A trigger event can be used to indicate the start or end of a scan pixel or scan line. In this case, the data stream can not be split into buffers that are processed concurrently, because events have to be parsed chronologically. Time-to-digital triggers could be used to track the clock drift between the scan unit and the detector, but this can at best form a software replica of a phase-locked loop that is better achieved on a hardware level. Additionally, a master clock allows for better expansion of the experimental setup and synchronization with other components. For this reason, large facilities like synchrotron facilities tend to rely on a master clock \cite{Serrano2009}.
    
    \subsection*{Clustering}
     An important consideration for event-driven operation is the difference between events per second and electrons per second, because of the activation of a cluster of neighboring pixels. For a frame-based detector, the cluster effect does not increase the data rate, and only affects the modulation transfer function (MTF). For an event-driven DED, the cluster effect increases the data rate and, as a result, it depends also on the chip's discriminator threshold. However, when TOA information is available, this can be corrected via a so-called declustering process, as has been demonstrated for sub-pixel precision imaging \cite{VanSchayck2020,Dimova2024}. For high-resolution 4D-STEM imaging, it was reported not to have a very significant effect on the reconstructed image \cite{Jannis2022}. It was also reported that this de-clustering can be performed in real-time \cite{Kuttruff2024}, which could provide a data rate reduction to the further processing pipeline.
     Histograms of the cluster size distribution for the experimental datasets used in this work are provided in the supplementary information. For the theoretical assessments made in the rest of this work, an average cluster size of 3 pixels will be assumed, which is a typical value for a 200 kV electron at a suitable threshold and sensor layer thickness \cite{Jannis2022,Dimova2024}.

    \subsection*{Saturation and coincidence losses}
\begin{figure}
        \centering
        \includegraphics[scale=0.75]{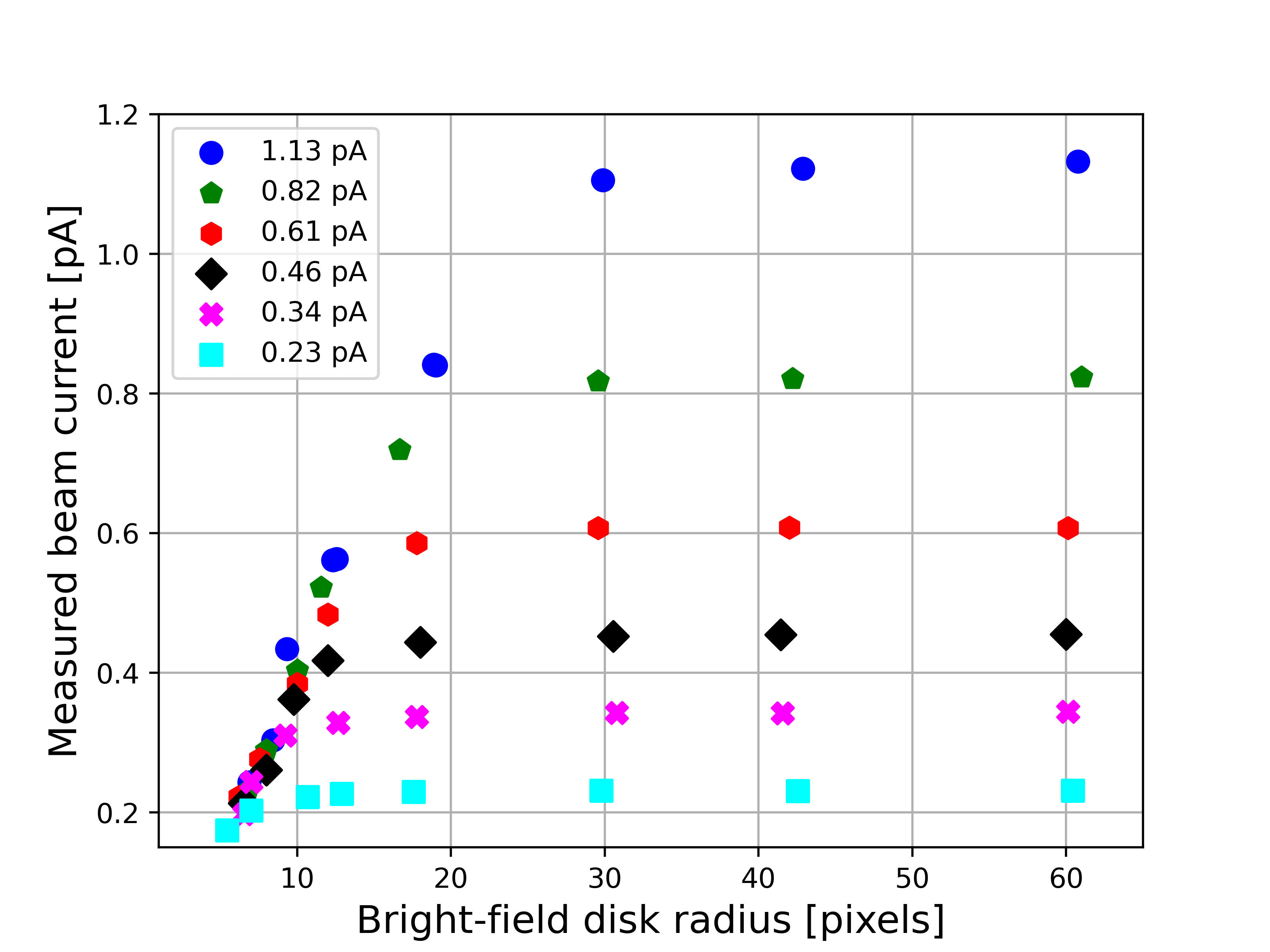}
        \caption{Measured beam current on a single Timepix3 chip as a function of the size of the bright field disk. The different colors represent measurements at different beam current, where the beam current for large disks equals the true value.}
        \label{fig:coincidencelosses}
    \end{figure}

    The main limitation encountered in performing 4D-STEM experiments using a Timepix3 DED is the upper limit on the electron beam current, resulting from the pixel dead time or bandwidth limitations of the readout system. Coincidence losses occur when an electron hits a pixel during its dead time, i.e. the period needed to process and transfer data from a previous event, resulting in the new event not being registered. The maximum hit rate of a single Timepix3 chip is around 80 million events per second, resulting in a maximum beam current of a few pA, depending on e.g. the acceleration voltage and discriminator threshold. As a result, event-driven 4D-STEM has mostly been applied for the low-dose investigation of beam-sensitive materials \cite{Hugenschmidt2024,Kavak2025,Schrenker2024}. All DED can suffer from saturation. For frame-based HPAD, the dominant form of saturation arises due to exceeding the dynamic range allowed by the bit depth of the pixel counter. Overflow of the dynamic range is usually clearly noticeable, but this is not necessarily the case for coincidence losses. Coincidence losses can cause significant errors to creep into quantitative analysis \cite{Hattne2023,Li2013b,Vlahakis2025}, for example, estimation of the dose used for a measurement. Coincidence losses arise from a saturation on a pixel level, but bandwidth limitations of the readout system can also cause saturation on a chip or entire detector level. The saturation on the pixel level, and to a lesser extent on the chip level, depends heavily on the illumination condition. The most important parameters are the beam current and the amount of pixels over which the bright field disk is spread. In figure \ref{fig:coincidencelosses}, the coincidence losses, for different combinations of beam current and bright field disk radius, are compared for a beam in vacuum on a single chip of an Amsterdam Scientific Instruments CheeTah T3 Quad. As was reported in ref. \cite{Jannis2022}, when the bright field disk covers a large area of the chip, coincidence losses can be neglected in the current regime allowed by the global bandwidth of the Timepix3. This condition is common for high-resolution imaging when a large convergence angle is used. However, as the pixel-wise current increases by localizing the beam in a smaller area, the measured counts can become severely underestimated. As discussed in the rest of this work, event-driven detection can also be of great use in nano-beam scanning diffraction experiments. However, as figure \ref{fig:coincidencelosses} depicts, the constraints for a quantitative interpretation are a lot more stringent here. Unless an extremely low beam current, e.g. of the order of 10$^{-1}$ pA, is used, a small bright field disk is almost guaranteed to have an underestimated number of counts. Diffracted beams can also suffer coincidence losses if their effective beam current, which is often orders of magnitude weaker than that of central beam, is sufficiently high. In our case, severe saturation of pixels reveals itself in the intensity profile of the disk, which is shown in supplementary figure 1.
    
    \subsection*{Processing}
    In terms of data processing, when the $\left(r_x,r_y,k_x,k_y\right)$ set is known for any electron in the measurement, its contribution to a reconstruction can often directly be computed without the need for a memory-expensive dense four-dimensional array. For a virtual detector reconstruction this direct contribution is trivial, being a counter increment at the real-space coordinate if the momentum-coordinate is inside a predefined range, corresponding to e.g. an annular mask. The contribution to the CoM vector map is trivial as well. Through the guided progressive reconstructive imaging (GPRI) framework of analytical ptychography, which will be explained in details in a separate publication, the contribution of an electron to an electrostatic potential measurement also becomes straightforward. Finally, since the contribution of each electron can be treated individually, there is no need to store information on events that have already been processed, resulting in minimal memory requirements.

\section*{Scaling laws}
    
    \subsection*{Data size}
        
        Due to the introduction of pixelated detectors into STEM workflows, the amount of data generated in typical experiments has seen an enormous increase and is now routinely in the regime of tens or hundreds of GB per single scan. The scaling laws describing the size of a 4D-STEM dataset as a function of the acquisition parameters are very different between the event-driven and the conventional frame-based mode. In the event-driven case, there is no longer an explicit dependency of the data size on the number of visited probe positions or the number of detector pixels. Instead, the data size only explicitly depends on the total number of electrons, so that beam current and dwell time become determining acquisition parameters.
        
\begin{figure}[h!]
            \centering
            \includegraphics[scale=0.5]{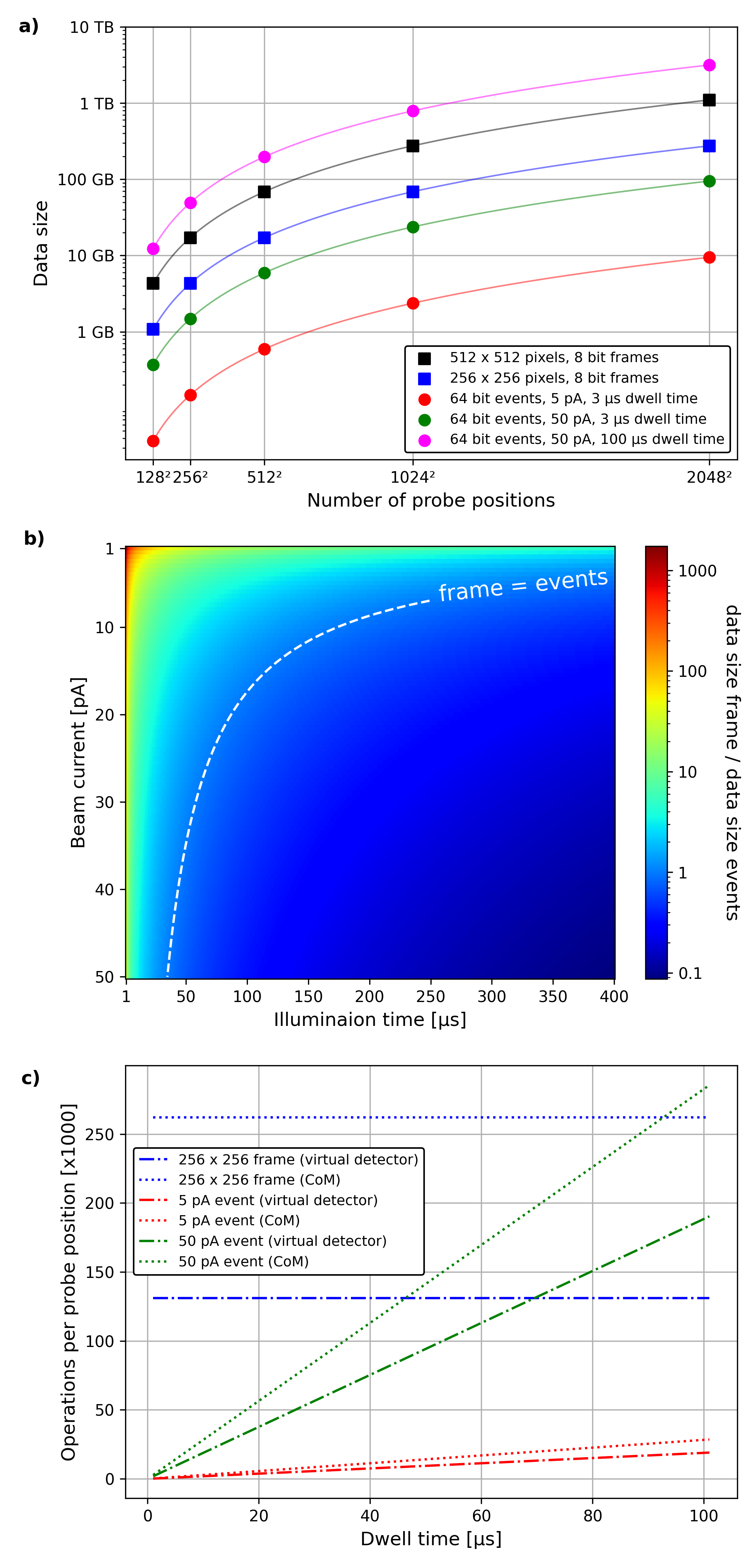}
            \caption{\textbf{a)} Data size of a 4D-STEM dataset as a function of number of probe positions, for different acquisition parameters, in the event-driven and frame-based scenarios.
            \textbf{b)} Data size ratio of a 64 bit event representation and an 8 bit frame representation for a detector of 512$\times$512 pixels, as a function of beam current and illumination time.
            \textbf{c)} Number of operations required to calculate a virtual detector count and CoM value for a single probe position, in the event-driven and frame-based scenarios. For all, an average cluster size of 3 pixels is assumed. It must be noted that not all scenarios considered here can actually be realized with current generation HPAD, such as a beam current of 50 pA with event-driven detectors or the fast frame rates with frame-based detectors.}
            \label{fig:data-rates}
        \end{figure}

        Figure \ref{fig:data-rates}a shows the evolution of the data size of a 4D-STEM dataset as a function of the number of probe positions in different scenarios. The explicit dependency on the number of recorded electrons gives not only a pronounced benefit in low-dose conditions, but also in scenarios that would not typically be considered low-dose. This is exemplified in figure \ref{fig:data-rates}b, where the data size ratio of a 64 bit event representation and an 8 bit frame representation are compared as a function of beam current and illumination time for a detector with 512$\times$512 pixels. In typical high-dose scenarios, the frame-based representation can achieve a benefit of about one order of magnitude in the considered range, whereas the benefit of the event-driven representation can reach more than three orders of magnitude in the considered range. For example, using a beam current of 50 pA and a dwell time of 30 $\mu$s would generally not be considered low-dose, but is still in the regime where the event representation is more efficient, given the configuration assumed for figure \ref{fig:data-rates}b. Figure \ref{fig:data-rates}c relates to the amount of computation required to process the data, which will be discussed further on.
        
\begin{figure*}[h]
            \centering
            \includegraphics[scale=0.45]{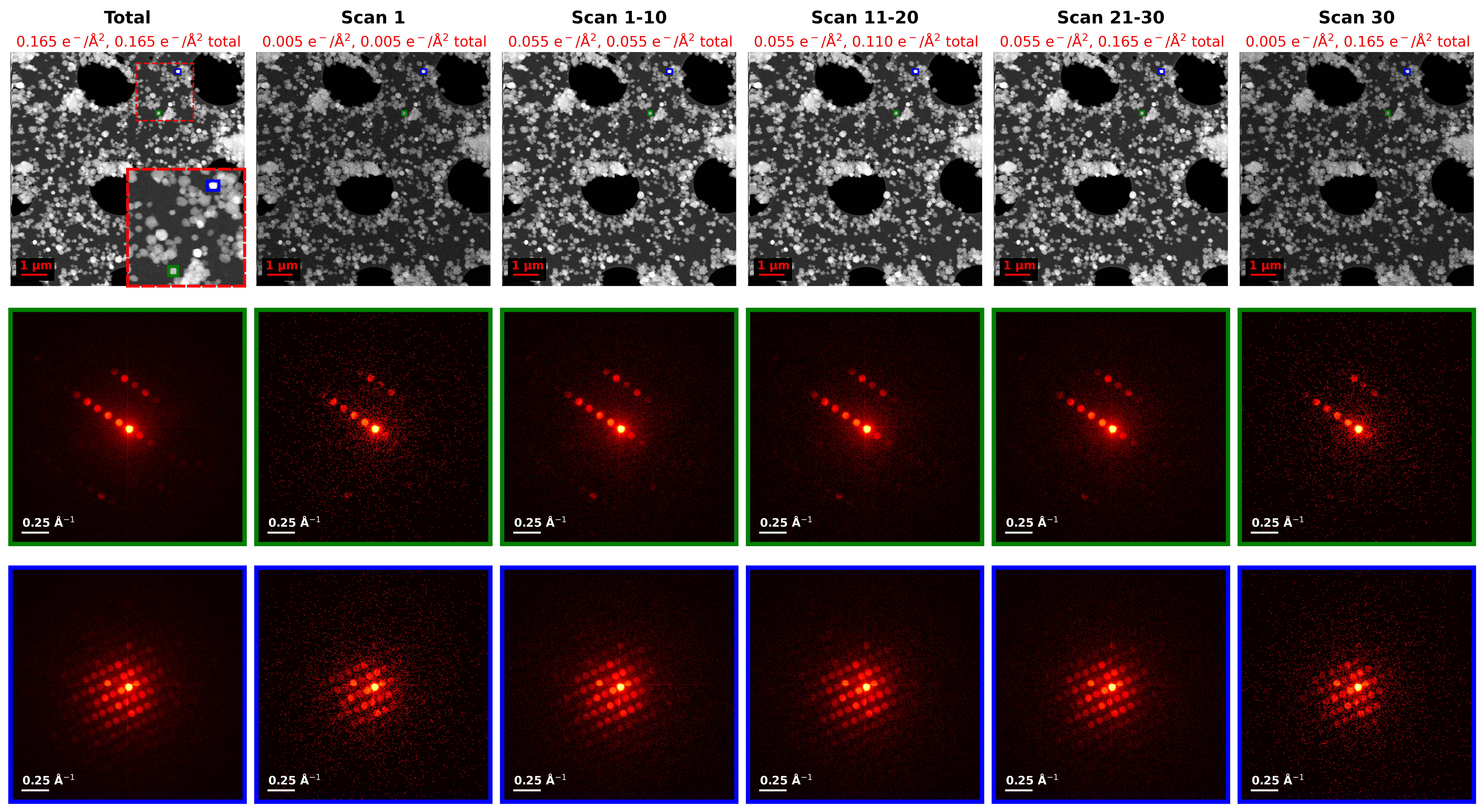}
            \caption{Demonstration of an event-driven multi-scan 4D-STEM measurement of a commercial ZIF-8 metal-organic framework, recorded using an Amsterdam Scientific Instruments CheeTah T3 Quad with 512$\times$512 pixels. The measurement consists of 30 scans with 2048$\times$2048 scan positions using a dwell time of 5 $\mu$s. A beam current of 0.4 pA is used to mitigate coincidence losses. The top row shows virtual ADF reconstructions computed directly from the event data. The bottom rows shows the diffraction patterns from the particles in the region of interest shown by the boxes in the images, extracted directly from the events. The total data size is approximately 34 GB, while an uncompressed 8 bit frame format would be approximately 33 TB.}
            \label{fig:multican}
        \end{figure*}

        It is noteworthy that the event-driven setup has mostly been demonstrated for high-resolution STEM imaging. To demonstrate that these alternative scaling laws can be beneficial in a broad range of applications, figure \ref{fig:multican} shows an example of a nano-beam 4D-STEM dataset of a commercial ZIF-8 metal-organic framework using 30 scans of 2048$\times$2048 positions using 5 $\mu$s dwell time, covering a field-of-view of approximately 10 $\mu$m. Thanks to the event representation, the total data size is approximately 34 GB, while an uncompressed 8 bit frame format would require approximately 33 TB. Further experimental details are provided in the supplementary information. Figure \ref{fig:multican} demonstrates how, thanks to the event mode of the detector, the resolution obtained in real and reciprocal space simultaneously is much closer to being directly limited by physics rather than practical constraints, while maintaining high speed and, as a result, low electron dose. Since the data size only scales with the number of counts, it does not depend on the fractionation of a measurement into multiple scans. For a frame-based detector, the splitting of a single scan with dwell time $\tau$ into $n$ faster scans using $\frac{\tau}{n}$, which in the first place is bound to the lower limit on the frame time, comes with an increase of the data size of a factor $n$. The ability to split a single scan into multiple high-speed scans brings insight into the dynamics of the sample, that would otherwise be hard to achieve, at no cost. Furthermore, it has been shown that for certain samples, a reduction of the dose rate, which can be achieved through such a fractionation into multiple scans, can significantly reduce the induced electron beam damage \cite{Velazco2022,Jannis2022a}. Additionally, the fractionation of a scan into multiple fast scans also allows to correct sample drift \cite{SAVITZKY2018}.
        
        Continuing, figure \ref{fig:multican} depicts the evolution of the spatially resolved diffraction information at dose intervals of approximately 0.005 e$^-/$\AA$^2$ per scan. No significant changes in diffraction intensity due to beam damage are observed, which is expected, as the dose applied here is far under the critical dose reported for ZIF-8 in the literature \cite{Liu2020z,Banerjee2024}. The ability to routinely record such datasets with large fields of view and high real- and reciprocal space resolution, at many small dose intervals, holds promise for applications like STEM serial electron diffraction \cite{Hogan-Lamarre2024} or 4D-STEM tomography \cite{Gallagher-Jones2020,Gholam2024a}.
    
    \subsection*{Data processing}
        
        Naturally, the benefits and scaling laws regarding data size have a very similar counterpart in the amount of computation required to extract information from a dataset. Figure \ref{fig:data-rates}c depicts the evolution of the number of numerical operations needed to extract a virtual detector count or CoM value from a diffraction pattern as a function of the illumination time. Except for very high dose scenarios, a significant benefit is obtained in the event-driven case. This is of high importance, as there is a growing need for workflows that provide the benefits of 4D-STEM live at the microscope, while the experiment is being performed. For high-resolution imaging, there have been several efforts in this regard, including for example for iterative ptychography \cite{Weber2024}, WDD \cite{Bangun2023}, iCoM \cite{Yu2022a}, SBI \cite{Strauch2021} and neural network-based reconstructions \cite{Friedrich2023}.
        
        This need for live processing of large datasets generated by 4D-STEM has created a trend toward the incorporation of high-performance computing (HPC) infrastructure into EM workflows \cite{Vescovi2020,Wang2022c,Mukherjee2022,Ercius2024,Pratiush2024a,Welborn2024}.
        Although in certain cases the need for such infrastructure might be inevitable, this work shows that, in many cases, efficient representation of electron data enables in-line processing using standard microscope control hardware—such as a desktop computer—with sufficient performance. HPC workflows do not only come with large costs and limited accessibility, but they potentially also have higher latency than local computing infrastructure, stemming from e.g. job scheduling and data transfer to and from shared storage. Such high latency is an important limitation for HPC when aiming for truly live processing, for example in using reconstructions from data streams in a search mode when finding a sample or optimizing the optics of the microscope. This is an important consideration as, for example, the benefits provided by dose-efficient 4D-STEM methods are strongly mitigated when the user is still limited to less appropriate methods like ADF imaging when setting up for acquisition, and is forced to choose between either damaging the sample anyway or working blindly. These considerations regarding the implications of HPC for 4D-STEM data processing, like cost, barrier to entry, accessibility and latency, are important to consider now, as they determine the future of these 4D-STEM methods and their integration in the wider field. Aiming for accessible, fast and efficient workflows increases the chances of 4D-STEM methods to establish themselves as routine and widespread tools for the investigation of e.g. life science specimens or other beam-sensitive materials.

    \subsection*{Ideal data representations}
        
        In the discussion above, the distinction between frame-based and event-driven detectors is made on the level of the chip application-specific integrated circuit (ASIC), as in the case of the Medipix3 and Timepix3 chips. However, the read-out mechanism of a DED does not necessarily have to be considered fixed to that of the ASIC, as it is possible to transform the initial data stream between different representations \cite{Datta2021,Guo2020,Pelz2022,Ercius2024}. This transformation could happen on a low hardware level, before the data stream reaches the user's processing computer, for example by performing run-length encoding using a field-programmable gate array (FPGA). Many of the properties of the DED will remain determined by the configuration of the ASIC, like the maximum beam current, the maximum speed, and the information content, e.g. the presence of TOA information. Consequently, this will determine the range of applications allowed by the DED, such as e.g. time-resolved measurements \cite{Jannis2021,Varkentina2022,Castioni2025,Borrelli2024,Haindl2023,Preimesberger2025}.The arguments presented above remain valid irrespective of the initial configuration of the DED, as these simply consider sparse representations of electron data rather than the more classical dense representation.
        
        Both the frame and event representation are available as the direct readout format of the current generation DED. However, even more efficient representations can be achieved after additional transformations, such as the true sparse representation or a lossless-compressed representation using compression methods such as e.g. Gzip \cite{deutsch1996,gailly1993}. The true sparse representation removes the redundancy of multiple events activating the same pixel, only storing active pixel indices and a pixel counter value. The data size of the frame and event representation does not depend on the distribution of counts, and could therefore be compared generically in the sections above, whereas the sparse and compressed representations do depend on the distribution of counts. In table \ref{tab:tableDS}, the 8 bit frame, 64 bit event, 8 bit sparse, Gzip-compressed frame and Gzip-compressed sparse representations are compared for 4 typical realizations of scanning diffraction patterns. Once again, it is clear how the event representation in many cases brings a significant size reduction compared to the full frame format. As expected, the true sparse representation brings an additional benefit compared to the event representation. For case \textbf{a} and \textbf{c}, where the counts per pattern are in the order of hundreds, the uncompressed sparse representation is the lightest. In case \textbf{b} and \textbf{d}, where the counts per pattern are in the order of several thousands, the compressed frame representation is the lightest. It is nevertheless important to stress here that the Gzip format cannot be processed directly. This comes with the disadvantage of needing a supplementary computational step to compress and decompress the data. An additional issue for the frame Gzip representation is that, upon processing, one has to fall back to the frame representation with its associated heavy computational and RAM requirements, as discussed earlier. Overall, lossless-compressed representations, such as e.g. Gzip or even more efficient compression frameworks \cite{Matinyan2023}, can be very suitable for archiving raw data that is not processed regularly, while the event and sparse representation provide the best trade-off between data size and direct accessibility.
        
        \begin{figure*}[h]
            \centering
            \includegraphics[scale=0.65]{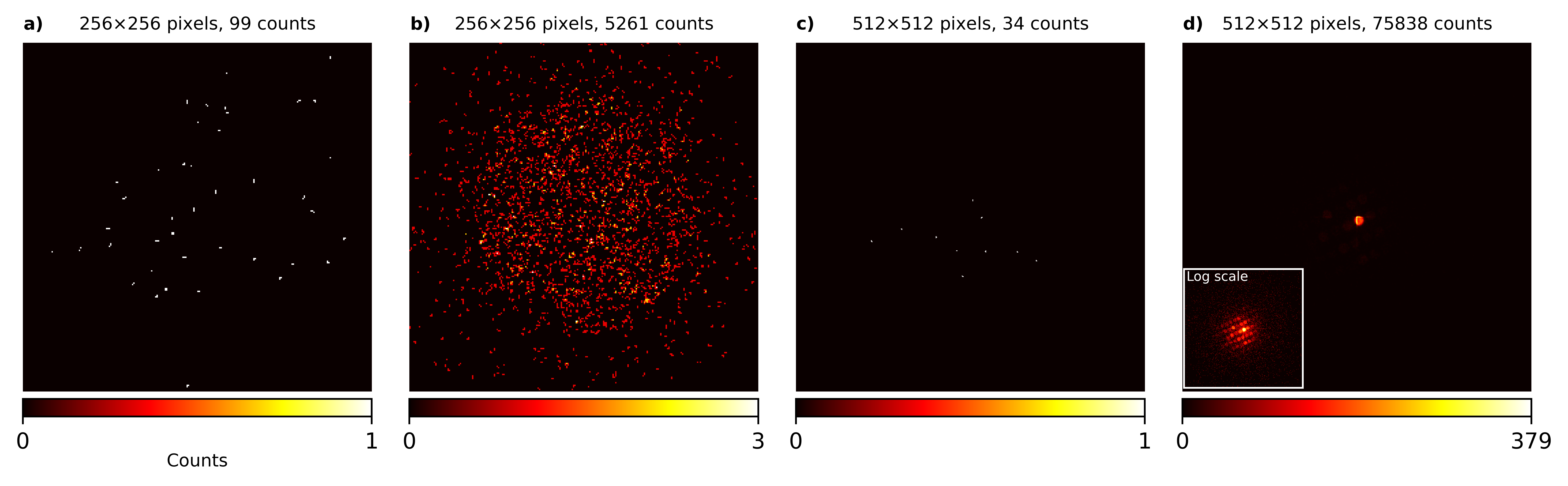}
            \caption{Examples of diffraction patterns with different illumination conditions, for which the data size in different representations is compared in table \ref{tab:tableDS}. Case \textbf{a} is a CBED pattern corresponding to a single probe position taken from the dataset of figure \ref{fig:benchmark}, with 6 $\mu$s dwell time at a beam current of approximately 1 pA. Case \textbf{b} is the position averaged CBED pattern of 50 scan points in the same dataset, which mimics the total amount of counts that would have been reached if a beam current of approximately 50 pA had been used. Case \textbf{c} is the diffraction pattern of a single probe position in the dataset used in figure \ref{fig:multican}. Finally, case \textbf{d} is the average diffraction pattern of the ROI in figure \ref{fig:multican}. For pattern \textbf{d}, an unsigned 8 bit integer representation would not provide the required dynamic range and information would thus have to be sacrificed.}
            \label{fig:compression}
        \end{figure*}
        
        \begin{table*}[h]
            \centering
            \begin{tabular}{|c|c|c|c|c|c|c|c|}
            \hline
            \textbf{Pattern}& \textbf{Pixels} & \textbf{Counts} & \multicolumn{3}{c|}{\textbf{Data Size} [KB]} & \multicolumn{2}{c|}{\textbf{Data Size Gzip} [KB]} \\
            \cline{4-8}
            & & & Frame & Event & Sparse & Frame & Sparse \\
            \hline
            \hline
            a & 256$^2$& 99 & 65.54 (100\%)& 0.79 (1.2\%)& \textbf{0.30 (0.5\%)}& 0.33 (0.5\%) & 0.38 (0.6\%) \\
            \hline
            b & 256$^2$ & 5261 & 65.54 (100\%) & 42.09 (64.2\%) & 14.74 (22.5\%)  & \textbf{4.40 (6.7\%)}  & 11.59 (17.7\%) \\
            \hline
            c & 512$^2$ & 34 & 262.14 (100\%) & 0.27 (0.1\%)& \textbf{0.17 (0.1\%)}& 0.45 (0.2\%) & 0.22 (0.1\%) \\
            \hline
            d & 512$^2$ & 75838 & 262.14 (100\%) & 606.70 (231.4\%) & 137.13 (52.3\%)& \textbf{7.23 (2.8\%)} & 50.74 (19.4\%)\\
            \hline
            \end{tabular}
            \caption{Comparison of the data size of the experimental diffraction patterns shown in figure \ref{fig:compression} for 8 bit frame, 64 bit event, 8 bit sparse, Gzip frame and Gzip sparse representations. The most data-efficient representation is marked in bold.}
            \label{tab:tableDS}
        \end{table*}
    
    \subsection*{Other sparse approaches}
    Introducing sparsity in a 4D-STEM measurement, either in real- or momentum space, can significantly alleviate the practical constraints. Sparsity can be introduced in real-space through sub-sampling strategies and inpainting \cite{Robinson2024,Hsu-Chih2024,Smith2025} which, in the Poisson noise-limited scenario that is now reached, cannot raise the amount of Fisher information above that of regular sampling \cite{VandenBroek2019a}. When considering the $\left(r_x,r_y,k_x,k_y\right)$ information of an electron in a 4D-STEM measurement, sparsity is not wanted in real-space (where the inpainting is performed) while it is both inherent and welcome in momentum space. Therefore, event-driven detection is a much more natural approach towards removing the practical constraints for low dose 4D-STEM compared to sparse scanning approaches.

\section*{The evenTem software framework for event-driven 4D-STEM}
    
    As discussed above, there can be many benefits associated to performing 4D-STEM in a sparse manner through the use of event-driven pipelines. However, to leverage these benefits, a software framework is required that is designed specifically with the intention of acquiring and processing 4D-STEM in a manner optimized for the event format. In particular, and as discussed above, synchronization routines in event-driven 4D-STEM can be quite distinct from those used in the more established frame-based setups. Naturally, the processing workflow also has to be set up with consideration given to the synchronization method applied at acquisition.
    
    To fulfill those needs, the evenTem framework was developed, as an extension of prior developments by Yu et. al. for real-time iCoM (riCoM) reconstructions \cite{Yu2022a}. Overall, evenTem aims at bridging the gap between the STEM community and event-driven workflows, as well as between the acquisition and processing side of 4D-STEM. It is based on a modular design that facilitates expansion to new hardware, in particular other pixelated detectors, and new processing methods. The framework furthermore supports acquisition and processing in dense modes using frame-based DED, though this is outside the scope of this manuscript.
    
    \subsection*{Acquisition}
        
        Currently, the event-driven DED for which acquisition control is already implemented include the Amsterdam Scientific Instruments CheeTah T3 and a custom prototype built from a commercial Advacam AdvaPIX T3. The frame-based Quantum Detectors MerlinEM \cite{krajnak2023,Nord2020} is also supported, which offers acquisition in a 1 bit mode, having interests for straightforward sparsification \cite{OLeary2020}. Regarding scan control, the framework currently supports the use of the Quantum Detectors scan engine, which provides possibilities for various triggering and synchronization functionalities including e.g. line triggering, pixel triggering, master clock generation and pattern scanning. These functionalities are aligned with the processing pipeline to enable seamless synchronization in many configurations. For the integration of the microscope column's hardware and functionalities, there currently is support for Thermo Fisher Scientific instruments employing the temscript Python wrapper \cite{Niermann2012}. Integration of JEOL systems is planned for future work. A graphical user interface (GUI) is available providing an interface to the centralized and synchronized control over all hardware components and live feedback of experimental output.
    
    \subsection*{Processing}
       Live processing of the raw data streams provided by the detector, encompassing e.g. the decoding of raw events to usable electron coordinates, requires processing rates of the order of GB/s. These performance requirements warrant the need for an efficient implementation in a high-performance language, which is in our case the C++ language. Another framework for the processing of event-driven STEM experiments, dedicated towards EELS, was developed by Auad et al. \cite{Auad2022} and is based on the Rust programming language.
        
        To facilitate the use of the reconstruction pipeline in the larger ecosystem of live experiment control or other processing workflows, the C++ functionality is wrapped into a Python module using Pybind11 \cite{Jakob2017}. The Python Application programming interface (API) simplifies integration into the GUI or into scripting workflows. Available processing methods working directly on event data streams, either from a socket connected to the detector or raw acquired files, include virtual detector reconstructions, CoM imaging, the riCoM algorithm, the extraction of the diffraction information from a region of interest and the declustering of events. Additionally, a converter is available that can transform raw event streams into larger-than-RAM Gzip-compressed dense 4D datasets, using the HDF5 data model \cite{The_HDF_Group_Hierarchical_Data_Format}. This is of interest for long term storage, as discussed above, and for the integration into existing processing workflows \cite{Paterson2020a}. The C++ processing tool, its Python API and its GUI are available open-source on github in the \href{https://github.com/EMAT-Jo/evenTem}{EMAT-Jo/evenTem} repository. Hardware interfacing components are not included as they depend on proprietary information from the vendors. The scripts to replicate the results in this work are available on Zenodo \cite{Annys2025}.

\section*{Benchmarking the event-driven processing pipeline}
    
    Due to the limited spatial extension of the electron probe in STEM experiments, it is appropriate to employ processing methods based on algorithms that truncate the spatial information carried by a single scan position to a certain kernel. For virtual detector STEM, this kernel is trivially limited to the pixel of the current scan point. Through the riCoM algorithm \cite{Yu2022a}, this concept was used as a means to integrate the CoM vector field. In the GPRI framework, it is applied to analytical ptychography reconstructions. There, the contribution of each available scattering vector is precomputed in the form of a library. This mapping of scattering events to an expected contribution has to be performed only once for a given experimental configuration. Within the measurement itself, a recorded event is processed simply by adding this contribution, i.e. a kernel of size R $\times$ R, to an ongoing reconstruction. As a result, only the rates at which electron events can be parsed, and kernels added, determine the processing speed. In the following, we will simply assess the performance of a processing pipeline that applies this kernel-additive approach for the live processing of an electron event stream. As mentioned above, more details on the GPRI framework and its theoretical foundations will be provided in another publication.
    
    \subsection*{Computational complexity of kernel-additive reconstructions}
        
\begin{figure}[h]
            \centering
            \includegraphics[scale=0.6]{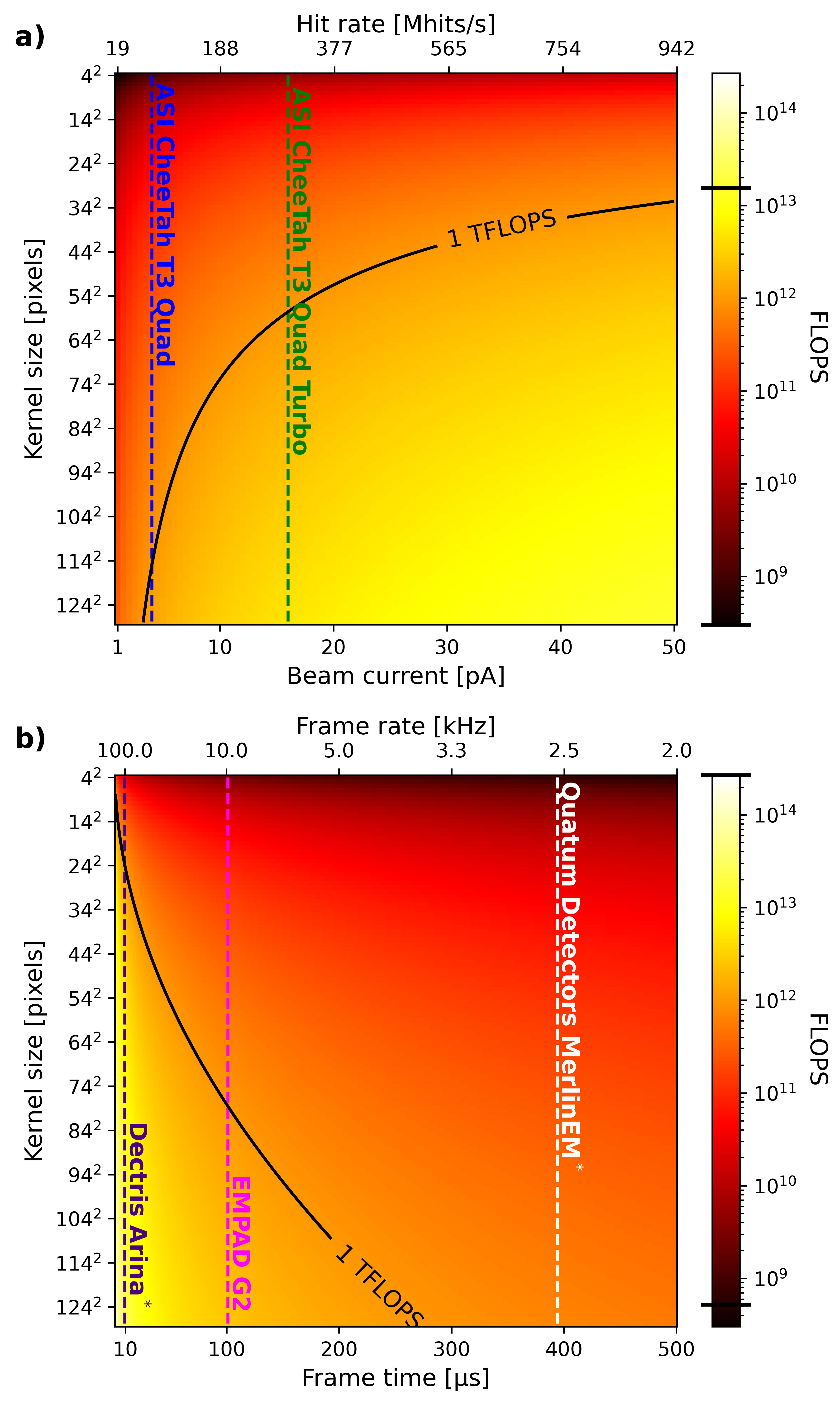}
            \caption{Assessment of the computational requirements of a kernel-additive method like GPRI as a function of kernel size and data rate. In \textbf{a}, the event hit rate and associated beam current (with a cluster size 3 pixels) is used, showing the maximum hit rate of current generation event-driven DED. In \textbf{b}, the frame rate and associated frame time is used, situating current generation DED \cite{Philipp2022,krajnak2023,Zambon2023}. To enable the comparison in \textbf{b}, a detector of 128 $\times$ 128 pixels is assumed, which is not the actual total number of pixels for the detectors marked with a star. The displayed FLOPS range is fixed to facilitate the comparison between \textbf{a} and \textbf{b}, with markings indicating the minimum and maximum value of each. The required FLOPS range from the order of GFLOPS to 100 TFLOPS, which is within the theoretical peak performance of a single current-generation consumer GPU.}
            \label{fig:FLOPS}
        \end{figure}

        In the case of the event-driven kernel-additive approach, the computational complexity of the reconstruction follows $\bigO(N \cdot R^2)$, with $N$ the number of events and $R^2$ the total amount of pixels in the kernel. As such, the number of floating point operations per second (FLOPS) that has to be made to live process a complete event-driven data stream, in the GPRI framework, can be directly estimated from the beam current and kernel size. A single-chip Timepix3 detector may provide a maximum of about 80 million events per second, which corresponds to a beam current of a few pA depending on the cluster size. A realistic size for a GPRI kernel is 64 $\times$ 64 pixels in the reconstruction window, and a single-precision floating-point representation is sufficient. In this scenario of a single precision 64 $\times$ 64 kernel, and assuming a beam current of 5 pA with an average cluster size of 3 pixels, the number of additions to be performed per second is just under 400 GFLOP. Going further, a more detailed evaluation of the required FLOPS, as a function of beam current or frame rate and kernel size, is presented in figure \ref{fig:FLOPS}. Standard graphical processing units (GPU) available in desktop computers, like the Nvidia RTX series, have a theoretical peak performance in the tens to hundreds of TFLOPS regime. While the FLOPS performance achieved in practice is quite significantly less than this theoretical value, it then remains clear that the computational requirements are within an achievable order of magnitude. More generally, this means that, in the GPRI framework, there is no requirement for advanced hardware, that would go beyond standard consumer hardware, to meet the computational requirements of live analytical ptychography reconstructions from DED data streams. This nevertheless still requires a performant implementation of the processing pipeline, that can sufficiently leverage the parallel nature of a GPU.
        
    \subsection*{Numerical demonstration}
        
        In this subsection, a demonstration is made of the live processing performance using a consumer desktop computer consisting of a 16 core AMD Ryzen threadripper pro 5995wx CPU and an Nvidia RTX4090 GPU. Although this is a high-end desktop configuration, it remains extremely modest in comparison to typical HPC infrastructure. For repeatability, the evaluations are based on raw recorded datasets stored on an NVMe solid-state disk. However, the results are also representative for raw data streams coming from a socket connected to the detector. For GPRI, GPU acceleration is achieved using a combination of Pytorch's LibTorch API \cite{Paszke2019} and custom CUDA kernels \cite{Nickolls2008}. Before making an assessment of the processing speeds of the ptychographic reconstructions using the GPRI framework, an evaluation is made of the performance of the underlying evenTem pipeline for parsing the raw packets to $\left(r_x,r_y,k_x,k_y\right)$ coordinates. To this end, a virtual ADF reconstruction is made using a single CPU thread for reading data and a single CPU thread for parsing the events. The riCoM method is also used for comparison, as it has already been shown to provide live dose-efficient reconstructions, based directly on the event stream.
        
\begin{figure*}[h]
            \centering
            \includegraphics[scale=0.65]{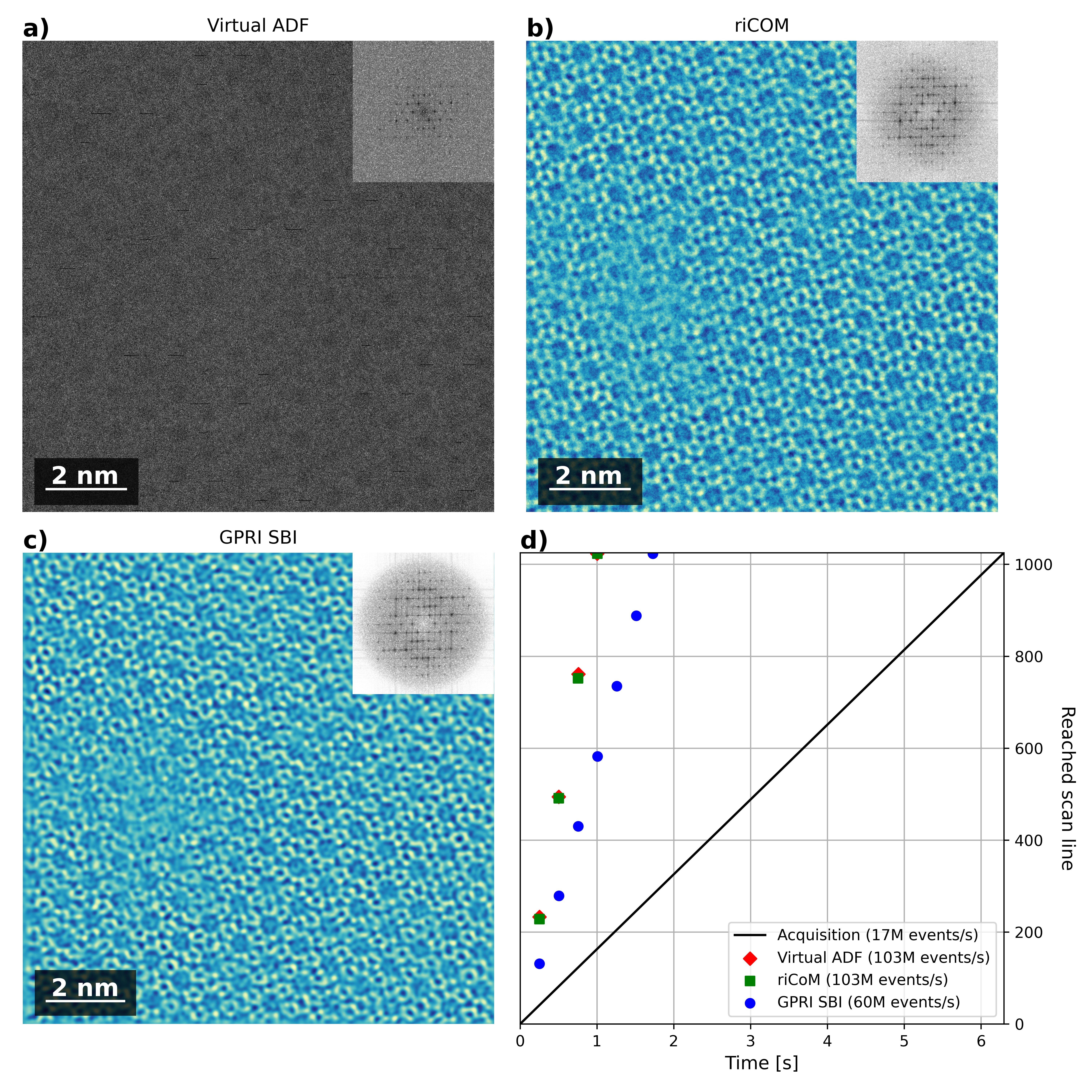}
            \caption{Panel \textbf{a)}, \textbf{b)} and \textbf{c)} respectively show a virtual ADF, riCoM and GPRI sideband integration reconstruction computed directly from the event-data. Panel \textbf{d)} compares the acquisition and processing rates by showing how many of the 1024 scan lines have been reached. All reconstruction rates exceed the acquisition rate for this dataset.}
            \label{fig:benchmark}
        \end{figure*}

        A dataset of silicalite-1 zeolite, used in ref. \cite{Jannis2022}, is employed which consists of 1024 $\times$ 1024 probe positions with a dwell time of 6 $\mu$s. The dataset was acquired using a beam current of approximately 1 pA corresponding to around 17 million events per second. Figure \ref{fig:benchmark}\textbf{a}-\ref{fig:benchmark}\textbf{c} depict the reconstructed images using respectively a virtual ADF detector, riCoM and GPRI SBI. The size of the integrating kernel for riCoM is 21 $\times$ 21 pixels and the size of the additive kernel for GPRI is 86 $\times$ 86 pixels. Figure \ref{fig:benchmark}\textbf{d} shows the acquisition rate and the processing rate of each method. Most importantly, each electron is processed directly and individually. As a consequence, the reconstructions can all be started simultaneously with the acquisition itself. The virtual detector and riCoM achieved a reconstruction rate exceeding 100 million events per second, resulting in a reconstruction time near 1 second. The GPRI process achieved a rate of 60 million events per second, resulting in a reconstruction time under 2 seconds. Furthermore, the processing rate for every method exceeds the acquisition rate, allowing truly live processing at the microscope, in the same conditions as when using a single pixel detector.
        
\section*{Discussion}
    
    The benefits of the 4D-STEM data acquisition and processing in an event-driven mode can be summarized as:
    \begin{itemize}
        \item When the time resolution of the detector is in the nanosecond or better regime, as in the case of Timepix3 and Timepix4, the detector does not impose the practical limit on the scan speed. Instead, this limit is imposed by response time of the scan system \cite{Jannis2022,Velazco2020}.
        \item Increasing the detector resolution comes without cost in data size. This is of high importance for nano-beam (scanning) diffraction experiments.
        \item Increasing the scan resolution (at a constant dose) comes without cost in data size. This enables large fields of view with sufficient sampling, which is of high importance for e.g. ptychography.
        \item Dividing a scan into multiple faster scans (at a constant dose) comes without cost in data size. This gives direct insight into dynamics due to e.g. beam damage or other stimuli. Such series provide a simple and direct way to assess critical dose post hoc and discard any data after the critical dose was exceeded.
        \item In many typical experimental scenarios, there is a strong reduction in the data size and the amount of computation required for information extraction compared to the conventional frame-based operation mode.
    \end{itemize}
    These benefits are mostly associated to the sparse representation of scattering data, and not necessarily to the specific operation mode of the detector. In principle, a frame-based HPAD capable of operating in the MHz frame rate regime combined with a transformation to a sparse representation on e.g. an FPGA would have many of the same attractions. Furthermore, the use of a pixel counter would make such a detector less susceptive to coincidence losses and bandwidth limitations.
    
    \subsection*{Outlook on Timepix4}
        
        The introduction of the Timepix4 chip \cite{Llopart2022} in the field of electron microscopy has been an exciting outlook, ever since the development of the first Timepix3-based EM workflows. The main improvements expected over this previous model are an improved time resolution, although for high-energy electrons this remains in question due to physical effects in the sensor layer \cite{Auad2024}, and an increased data rate, reducing the constraint on the maximum beam current. However, the effect of coincidence losses will still have to be assessed. Additionally, the Timepix4 chip will have improved frame-mode capabilities compared to the Timepix3 chip, which can be beneficial in experimental scenarios involving high doses or beam currents.
        
        Overall, this means that Timepix4-based detectors hold promises to become an ideal all-around solution for STEM, thus reinforcing the need to drive the integration of event-driven operation into electron microscopy workflows. The first experimental setups integrating the Timepix4 chip have already been realized \cite{Dimova2024}, although the maximum data rate performance has not yet been reported. From the assessments presented in this work, it is clear that the event format still provides numerous benefits over a dense format in many of the scenarios enabled by Timepix4. When utilizing its full potential bandwidth, Timepix4 would allow event rates far exceeding those generated by e.g. a typical 50 pA electron beam. To keep up with such an increased data rate, the presented processing pipeline will have to allow further parallelization. The current workflow uses a single CPU thread for data reading and a single CPU thread for event parsing. If events can be parsed independently of each other, an expansion to multiple CPU threads or a GPU should be straightforward. As discussed above, when synchronization is ensured by the use of a master clock, events can indeed be treated independently. A tenfold increase of the current processing rate, e.g. achieved through parallelization, would already allow live processing of the data stream resulting from a typical 50 pA electron beam.

\section*{Conclusion}
    
    As the frame rates of frame-based DED approach the $\mu$s frame time regime often desired for STEM experiments, they become increasingly inefficient in terms of data size and computational requirements. On the other hand, event-driven DED, such as those based on the Timepix3, offer a promising solution by operating in both sparse and continuous modes.
    
    In this work, a framework was introduced for the acquisition and live processing of 4D-STEM using fully sparse pipelines. The scaling laws of data and computation in event-driven and frame-based setups were compared, and it was shown how sparse pipelines can bring significant benefits in a broad range of applications, including scan resolution, field-of-view, detector resolution, acquisition and processing speed, minimum dose and dose-fractionation.
    
    Generally, many practical constraints associated with 4D-STEM can be effectively alleviated through event-driven workflows. Nevertheless, for the Timepix3 chip, the primary limitation is saturation, which imposes constraints on the maximum beam current. It was demonstrated how completely sparse pipelines can permit live processing and image formation, using a standard consumer computer and thus avoiding the need to introduce HPC infrastructure into electron microscopy workflows. Truly live processing, including for advanced methods like analytical ptychography, was demonstrated at rates at which usual single-pixel experiments like ADF-STEM are performed. This closes the gap to routinely driving experiments based on 4D-STEM information, instead of data treatment and information extraction being an afterthought or tedious post-processing step.

\section*{Acknowledgments}

    A.A. and J.V. acknowledge funding from the European Union under grant agreement no. 101094299 (IMPRESS). H.L.L.R. and J.V. acknowledge funding from the Horizon 2020 research and innovation programme (European Union), under grant agreement No 101017720 (FET-Proactive EBEAM). Views and opinions expressed are however those of the authors only and do not necessarily reflect those of the European Union or the European Research Executive Agency (REA). Neither the European Union nor the granting authority can be held responsible for them. S.G. , J.H. and J.V. acknowledge funding from an SBO FWO national project under grant agreement n. S000121N (AutomatED). S.G. and J.H. acknowledge funding from FWO grant G069925N.

\section*{Declaration of conflicts of interest}
    The authors declare no conflicts of interest.

\bibliographystyle{elsarticle-num}
\bibliography{biblio}

\end{document}